\documentclass[12pt]{iopart}
\usepackage{amssymb,graphicx,epsfig,iopams}
\begin{document}

\title[Influence of local structure on magnetic properties]{Influence of local structure on magnetic properties of Layered Cobaltites PrBaCo$_{2}$O$_{5+\delta}$, $\delta >$ 0.5}
\author{S. Ganorkar$ ^{1} $, K. R. Priolkar$^{1}$\footnote[1]{Author to whom correspondence should be addressed}, P. R. Sarode$ ^{1} $, A. Banerjee$ ^{2} $ and R. Rawat$ ^{2} $ and S. Emura$ ^{3} $ }
\address{$^{1}$Department of Physics, Goa University, Taleigao Plateau, Goa 403 206 India}
\address{$^{2}$UGC-DAE Consortium for Scientific Research, University Campus, Khandwa Road, Indore 452 001 India}
\address{$^{3}$Institute of Scientific and Industrial Research, Osaka University - 8-1 Mihogaoka, Ibaraki, Osaka 567-0047, Japan}

\ead{krp@unigoa.ac.in}
\date{\today}

\begin{abstract}
The effect of local structure on the magnetic and transport properties of the layered perovskites has been
investigated. The samples PrBaCo$_{2}$O$_{5+\delta}$, ($\delta$ =  0.80 and 0.67) though crystallize in the same
112 type tetragonal structure have different magnetic ground states. Extended X-ray Absorption Fine Structure
(EXAFS) spectroscopy has been employed to explain the changes in magnetic interactions based on the rotation  and
tilting of CoO polyhedra in these oxygen rich double perovskites.

\end{abstract}

\pacs{72.10.Di; 78.30.-j; 87.64.Je; 61.05.cj}
\vspace{2pc}
\noindent{\it Keywords}: Double perovskites, Magnetization, EXAFS, XANES

\submitto{\JPCM}

\maketitle

\section{Introduction}
Partial substitution of alkaline-earth metals for rare-earth metals in RCoO$_3$ (R - rare-earth element) and
their subsequent ordering of results in double layered perovskites. One such Ba substituted layered perovskites
are RBaCo$_{2}$O$_{5+\delta}$, ($\leq \delta \leq$ 1). The ionic size difference between Ba and R and the
resultant valance of Co due to oxygen non-stoichiometry make cobaltites more complex to study. They are of very
special interest for their unique properties like metal insulator transition (MIT), spin state transition, charge
ordering and giant magnetoresistance (MR)\cite{mai,mar,mor,res,vog,sua,tro}. These properties have their origin
in the strong interplay between crystallographic, magnetic and transport properties. A property that
distinguishes the cobaltites from other transition metal perovskites is the crystal field splitting of the
\textit{3d} energy level of the Co ion. It is of the same order of magnitude as the Hund's rule intra-atomic
exchange energy. Spin state transition can therefore be easily provoked by varying temperature, pressure,
magnetic field or chemical composition (oxygen content or rare earth ion) of the material. These spin state
transitions are believed to be responsible for their unique properties. The transport properties of the layered
cobaltites are found to be very sensitive due to the coexistence of ferromagnetic (FM) and antiferromagnetic
(AFM) state. A giant negative magnetoresistance (MR) (of about 10$\%$) coinciding with FM-AFM transition has been
observed in GdBaCo$_{2}$O$_{5.5}$ \cite{mar}. Whereas LaBaCo$_{2}$O$_{5.5}$ exhibits a strongly irreversible MR
near FM-AFM transition \cite{kun}.

Oxygen non-stoichiometry is one of the important aspect in these layered cobaltites. Tailoring of oxygen content
realizes in different valance states of cobalt ion (2+, 3+, 4+) and different local environments, both of which
play a crucial role in their magnetic properties \cite{str,yur,cfro}. The structure of Co based double
perovskites also depends on the oxygen content ($5+\delta$). In general, the lower and higher oxygen content
compounds crystallize in tetragonal structure with 112 type unit cell while the intermediate members crystallize
in orthorhombic 122 type unit cell \cite{yur,cfro}. The magnetic properties are found to be correlated to oxygen
content and crystal structure. While the compounds with lower $\delta$ have AFM ground state, FM dominates in
compounds richer in oxygen. When $\delta$ = 1, the  FM order is associated with metallic behavior due to double
exchange interaction \cite{ueda}. With the decrease in oxygen content, a transition from metallic to insulating
state occurs in the paramagnetic state and further as the number of trivalent Co ions increase FM order changes
to AFM ordering. Neutron diffraction studies on PrBaCo$_{2}$O$_{5.75}$ reveal a paramagnetic to FM and an
incomplete FM to AFM transition\cite{fr}. Furthermore, in these oxygen rich compounds, the Co$^{3+}$ ions in
octahedral coordination are found to be in HS state as compared to LS state in PrBaCo$_2$O$_{5.5}$ \cite{fron}.
It is therefore not clear if the HS to LS spin state transition of octahedral Co$^{3+}$ ions is responsible for
the transition from FM to AFM  ground state. Or it is the delocalization of pd$\sigma$ holes that are responsible
for the AFM to FM transition in these double perovskites \cite{wu}. Our recent studies have shown that local
structure around Co ions plays an important role deciding the magnetic properties of these hole doped cobaltites
\cite{shr}. Hence it would be worthwhile to explore the local structure around Co ions in these layered
cobaltites and correlate them with their magnetic properties. In particular, it would be interesting to examine
the changes in local structure of Co ions as the ground changes from FM to AFM.

Extended X-ray absorption Fine Structure (EXAFS) spectroscopy is known to be a valuable tool to probe the local
structure. EXAFS has been effectively used to understand the local structural distortions and the spin state of
Co ion in the parent RCoO$_{3}$ compounds \cite{pan,wudd,ber,pa,chang,haas,skp}. In comparison to RCoO$ _{3} $,
the oxygen cage around Co ion is more distorted in RBaCo$_{2}$O$_{5+\delta}$ \cite{kha,pom}. It is suspected that
the deformation of polyhedra modify the hybridization between outer orbits of cation (R,Ba,Co) and $2p$ orbitals
of oxygen ion which affect electric and magnetic properties of double perovskites \cite{hid}. Furthermore, the
difference between the ionic radii of R and Ba along with deformed polyhedra will affect the crystal field
splitting of the $3d$ band \cite{jarry}. Herein we report EXAFS spectroscopy investigations on two oxygen rich
PrBaCo$_{2}$O$_{5+\delta}$, ($\delta$ = 0.80 and 0.67) compounds crystallizing in the same 112 type structure but
having different magnetic ground states. While FM dominates in compound with $\delta$ = 0.8, the other one
($\delta$ = 0.67) has predominantly AFM ground state.

\section{Experimental}
The polycrystalline  samples of PrBaCo$_{2}$O$_{5+\delta}$ were prepared by sol-gel method. Stoichiometric
amounts of  Pr$_{6}$O$_{11}$, BaCO$_{3}$ and Co(NO$_{2}$)$_{3}$.6H$_{2}$O were dissolved in nitric acid. Citric
acid was added to the above solution as a complexing agent. The solution was then heated at 353K to form a gel
which was subsequently dried at 433K to remove the solvent. This precursor was ground, pelletized and heated at
1073K for 4 hours followed by annealing at 1323 K for 24 hours and slow cooling at the rate of 1$^\circ$/min to
room temperature to get the required sample. This as synthesized sample had an oxygen stoichiometry of $ \delta =
0.80 \pm 0.01$ as determined by iodometric titration \cite{con}. To obtain a sample with lower oxygen content,
the above as prepared sample was further annealed in Argon atmosphere at 673K for 36 hours and quenched in ice
water. The oxygen content as determined by iodometric titration was found to be $\delta = 0.67 \pm 0.01$. X-ray
diffraction (XRD) data were recorded using Rigaku D-Max IIC X-ray diffractometer  in the range of $ 20^\circ \le
2\theta \le 80^\circ$ with Cu K$_\alpha$ radiation to ascertain phase formation. The diffraction patterns were
Rietveld refined using FULLPROF suite and structural parameters were obtained. Resistivity measurements were
performed in the temperature range 325K-10K using four probe method. Magnetization measurements were carried out
as a function of temperature using a Quantum Design SQUID magnetometer in an applied field of 1000Oe and in the
temperature range of 10K to 300K. The sample was initially cooled from room temperature to the lowest temperature
(10K) in zero applied field. Magnetization was recorded while warming under an applied field (zero-field cooled
(ZFC)) and subsequent cooling (field-cooled cooling (FCC)) and warming (field-cooled warming (FCW)) cycles. The
magnetotransport studies were carried out in longitudinal geometry using OXFORD Spectromag 10T magnet. Resistance
as a function of temperature R(T) was measured in the range 10K - 300K in 0T and 8T during warming and cooling
cycles. Isothermal magneto-resistance MR was measured at selected temperatures by ramping the field in the range
$\pm$8T. The local structure around Co ion was investigated by measuring X-ray absorption fine structure (XAFS)
at the Co K edge at various temperatures between 325K to 20K. These experiments were carried out in transmission
using BL 7C at Photon Factory Japan. The absorbers were prepared by sprinkling finely powdered sample on a scotch
tape and stacking such layers to obtain optimum thickness such that $\Delta\mu t < 1$.

\section{Results}
\begin{figure}
\centering
\includegraphics[width=\columnwidth]{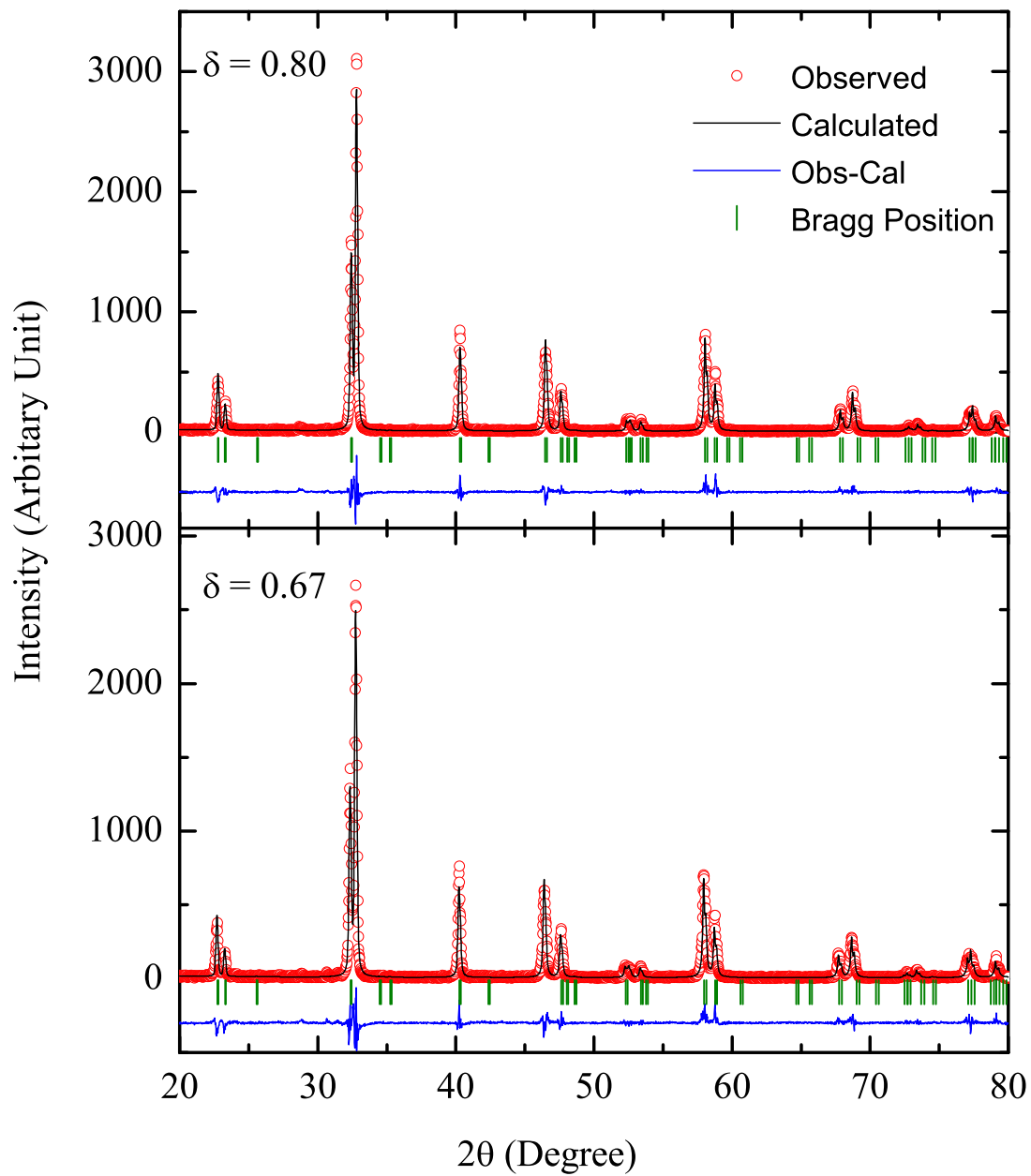}
\caption{\label{XRD} Rietveld refined XRD patterns of PrBaCo$_{2} $O$ _{5.5}$. Circles represent experimental
data, continuous line through the data is the fitted curve, vertical lines indicate Bragg positions and the
difference pattern is shown at the bottom as solid line.}
\end{figure}

Rietveld refinement of room temperature XRD patterns of PrBaCo$_2$O$_{5+\delta}$ for $\delta$ = 0.67 and 0.80
confirm the formation of single phase samples with 112 type tetragonal unit cell belonging to P4/mmm space group.
As can be seen from Figure \ref{XRD} both compounds show presence of very minor impurity phases ($\sim$ 1\%) with
peaks around 2$\theta \sim 29^\circ$ and 31$^\circ$ which can be ascribed to unreacted Pr-oxides (Pr$_2$O$_3$,
Pr$_6$O$_{11}$). The lattice parameters obtained from Rietveld refinement are given in Table \ref{lattice}. The
parameters obtained are in good agreement with the neutron diffraction results on PrBaCo$_2$O$_{5.75}$
\cite{cfro,fr}. However, high resolution synchrotron XRD measurements on similar composition reveal that the
structure belongs to orthorhombic unit cell with Pmmm space group \cite{gac}. In the present case no splitting of
(200) reflection as expected for an orthorhombic structure was seen. It must be noted here that a structural
crossover from tetragonal to orthorhombic unit cell is reported for oxygen concentration around $\delta$ = 0.6.
\cite{cfro,shr}.

\begin{table}[h]
\caption{\label{lattice}Lattice parameters for PrBaCo$_2$O$_{5+\delta}$.}
\footnotesize\rm
\begin{tabular*}{\textwidth}{@{}l*{15}{@{\extracolsep{0pt plus12pt}}l}}
\br
$\delta$&a(\AA)&b(\AA)&c(\AA)&V(\AA$^3$)\\
\mr
0.80&3.9045(1)&3.9045(1)&7.6355(2)&116.410(7)\\
0.67&3.9085(7)&3.9085(7)&7.6311(2)&116.579(5)\\
\br
\end{tabular*}
\end{table}

\begin{figure}
\centering
\includegraphics[width=\columnwidth]{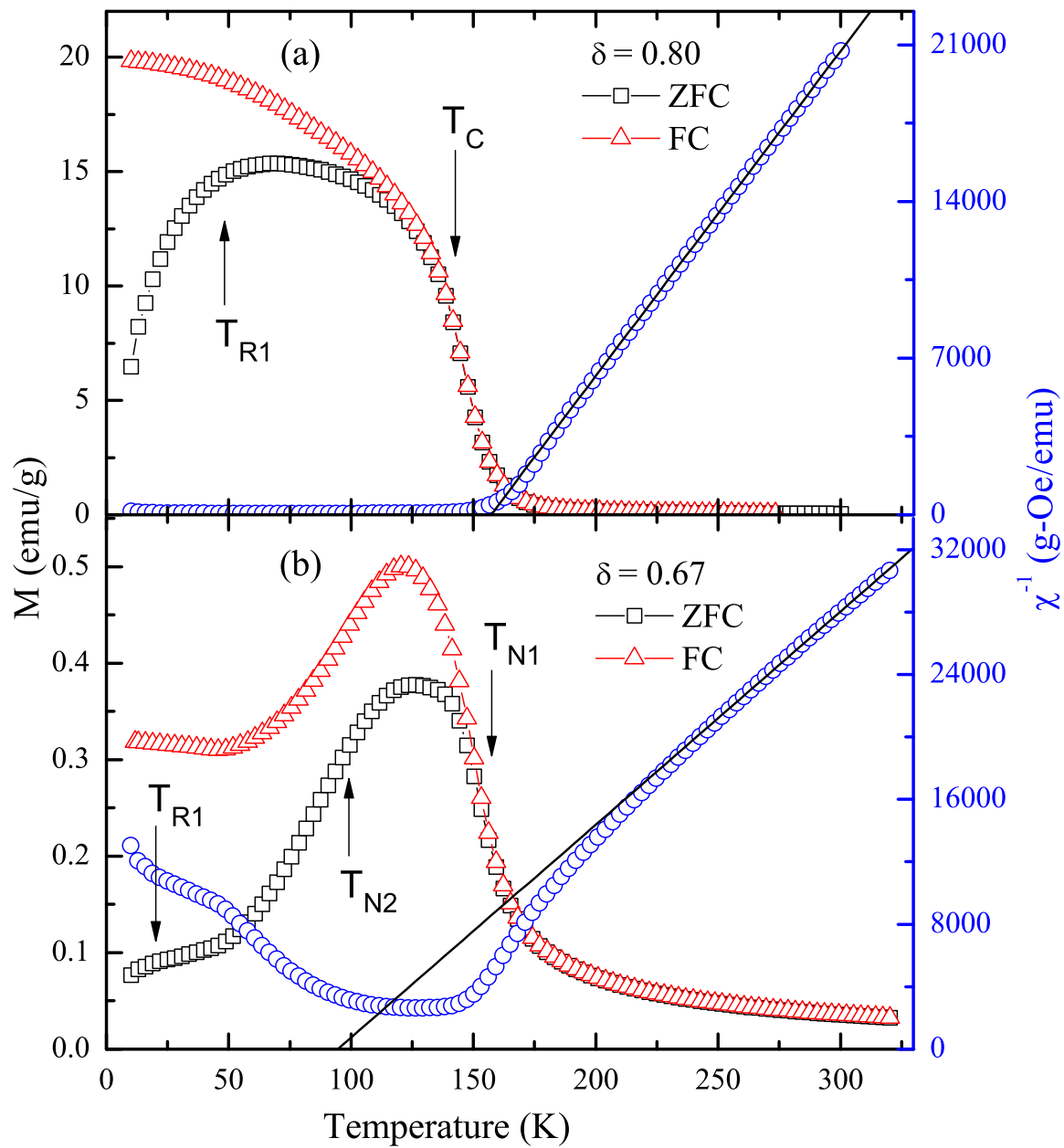}
\caption{\label{MT} Temperature dependent magnetization M(T)in the field of 1000 Oe for ZFC and FC cycle
indicated by open squares and triangles respectively. The magnetic transitions are indicated by arrows. 1/$ \chi $ Vs Temperature is denoted by circles along with
Curie-Weiss fit indicated by solid line. (a)$ \delta $ = 0.80 (b)$ \delta $ = 0.67}
\end{figure}

Temperature dependant magnetization M(T) measured at 1000 Oe during ZFC, FCC and FCW cycles for the two samples
is presented in Figure \ref{MT}. In case of $\delta$ = 0.80, a clear PM to FM transition can be seen at T$_{C}$ =
147K. The bifurcation seen between ZFC and FC curves at lower temperatures (T $\sim$ 120K) indicates a presence
of competing magnetic interactions. The general behaviour of magnetization is similar to that reported for
PrBaCo$2$O$_{5.75}$ \cite{fr}.

In case of $\delta$ = 0.67,  a small rise in magnetization, akin to a PM to FM transition is noticed at 150K.
However, the small value of magnetic moment and the subsequent sharp decrease in magnetization at T = 100K
prompts us to ascribe both these transitions at T$_{N1}$ = 150K and T$_{N2}$ = 100K to an AFM ground state. In
PrBaCo$_2$O$_{5.5}$, where magnetization has similar behavior, neutron diffraction studies have reported
antiferromagnetic spin alignment \cite{fron}.

One more transition, T$ _{R1} $ is observed at 50K and 37K for $\delta$ = 0.80 and $\delta$ = 0.67 respectively
which are more clearly shown in Ref. \cite{shr} and indicated by arrows in Fig. \ref{MT}. In literature these
have been attributed to reorientation of magnetic ordering due to structural transition \cite{jarry}.

Magnetic susceptibility $\chi = M/H$ calculated from magnetization also reveals important clues about magnetic
ordering in these two compounds. In the case of PrBaCo$_2$O$_{5.80}$, a plot of $\chi^{-1}$ vs T exhibits a
Curie-Weiss (linear) behavior almost down to its $T_C$. The values of paramagnetic Curie temperature, $\theta_p$
and effective magnetic moment $\mu_{eff}$ are 150K and 5.34 $\mu_B/f.u.$ respectively and are in good agreement
with those reported for PrBaCo$_2$O$_{5.75}$ \cite{fr}.

Deviation from Curie-Weiss fit are noted at temperature as high as 250K in case of PrBaCo$_2$O$_{5.67}$. The
deviation and the behavior of susceptibility can be ascribed to presence of weak antiferromagnetic correlations
well above its ordering temperature. Though $\theta_p$ obtained from Curie-Weiss fit is positive, there is a
decrease in its value from 150K for $ \delta $ = 0.80 to 92K for $\delta$ = 0.67. Canting of magnetic spins due
to different local environment of Co ions and/or their spin states could be the reason for positive values of
$\theta_p$. Neutron diffraction studies have indicated AFM ordering at T$_N$ = 220K in PrBaCo$_2$O$_{5.5}$
\cite{fron}. This ordering is a result of interaction between IS Co$^{3+}$ ions in square pyramidal coordination.
A possibility Co$^{3+}$ - O - Co$^{3+}$ pairs giving rise to weak antiferromagnetic correlations in
PrBaCo$_2$O$_{5.67}$, well above its magnetic ordering temperature cannot be ruled out. The value of $\mu_{eff}$
obtained from Curie-Weiss fit is 5.41$\mu_B/f.u.$


Based on the similarities of $\mu_{eff}$ values and their agreement with that of reported for
PrBaCo$_2$O$_{5.75}$ prompts us to assign LS state for Co$^{4+}$ ions and HS/IS state for Co$^{3+}$. It may be
also noted that a simple calculation of magnetic moment of Co ions assuming all Co$^{4+}$  to be in LS state and
octahedral Co$^{3+}$ ions in HS while square pyramidal Co$^{3+}$ in IS state gives a value of 5.47$\mu_B/f.u.$
which is in good agreement with that obtained from Curie Weiss fit. With a decrease in $\delta$, the
concentration of Co$^{3+}$ ions will increase giving rise to a higher possibility of Co$^{3+}$ - O - Co$^{3+}$
pairs. Clusters of square pyramidal Co$^{3+}$ ions can influence antiferromagnetic interactions at higher
temperatures.

\begin{figure}
\centering
\includegraphics[width=\columnwidth]{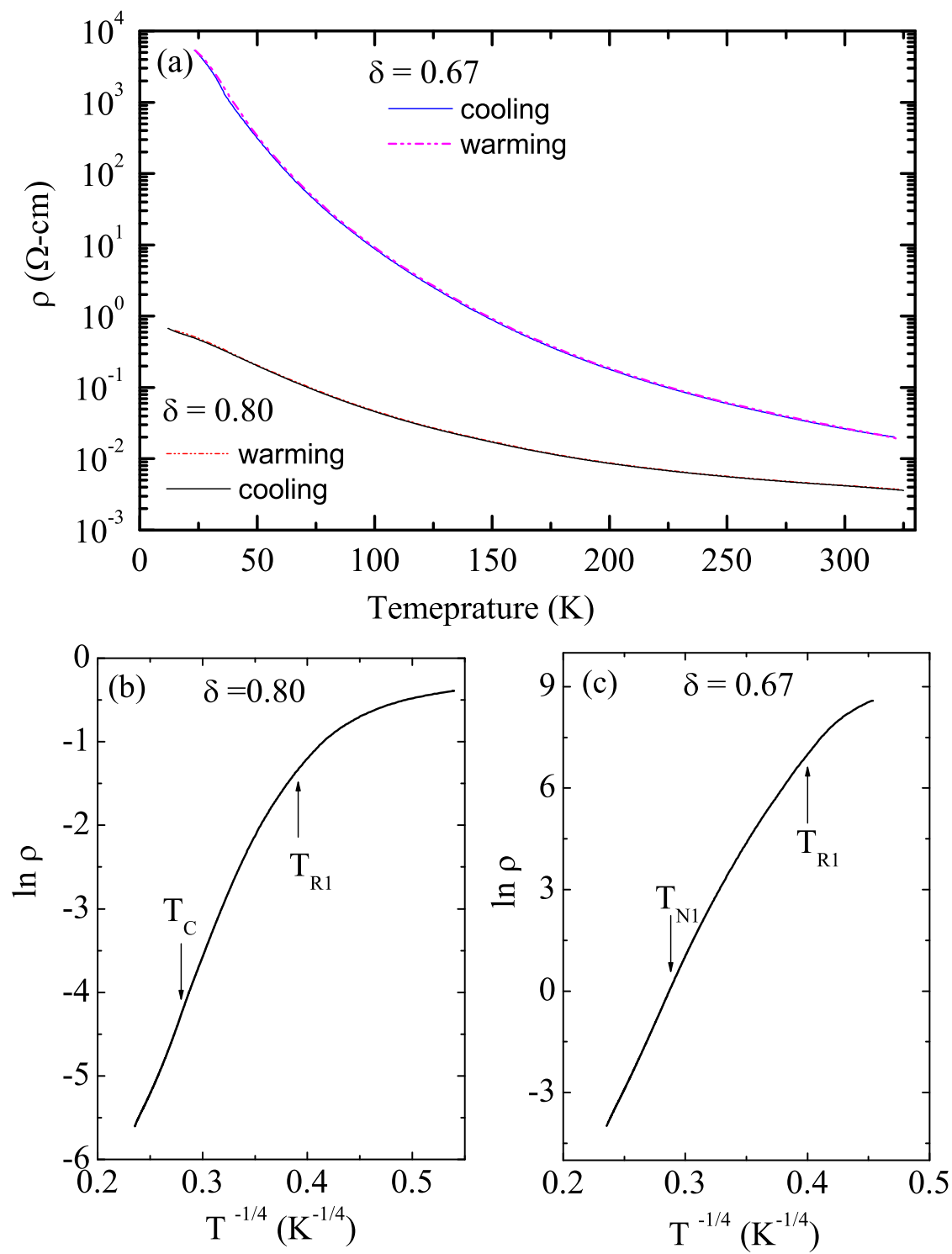}
\caption{\label{RES} (a) Plot of resistivity with temperature during cooling and warming cycles. (b) Logarithm of the resistivity verses T$ ^{-1/4} $ plot with arrows indicating different transitions.}
\end{figure}

Electrical resistivity exhibits semiconducting behavior with negligible hysteresis in the entire temperature
range (10K - 330K) for the two compounds, $\delta$ = 0.8 and 0.67 (Figure \ref{RES}(a)).  A comparison of
resistivity curves shows that resistivity as well as the change in resistivity are much higher for
PrBaCo$_2$O$_{5.67}$. For $\delta$ = 0.67, the resistance changes by almost 5 orders of magnitude from $\sim$
10$^{-1}$Ohm-cm at 300K to $\sim 5 \times 10^{3}$Ohm-cm at 25K while in case of $\delta$ = 0.8, it changes from
$\sim$ 10$^{-3}$ Ohm-cm to 10$^{-1}$ Ohm-cm in 300K $\ge$ T $\ge$ 25K. A plot of $\ln \rho$ v/s T$^{-1/4}$
presented in Figure \ref{RES}(b) and (c) exhibits a linear variation for a large temperature range indicating
variable range hoping (VRH) to be the dominant mechanism of conduction in both the compounds. However, the
hopping temperature is much higher for $\delta$ = 0.67  as compared to that in $\delta$ = 0.8.

\begin{figure}
\centering
\includegraphics[width=\columnwidth]{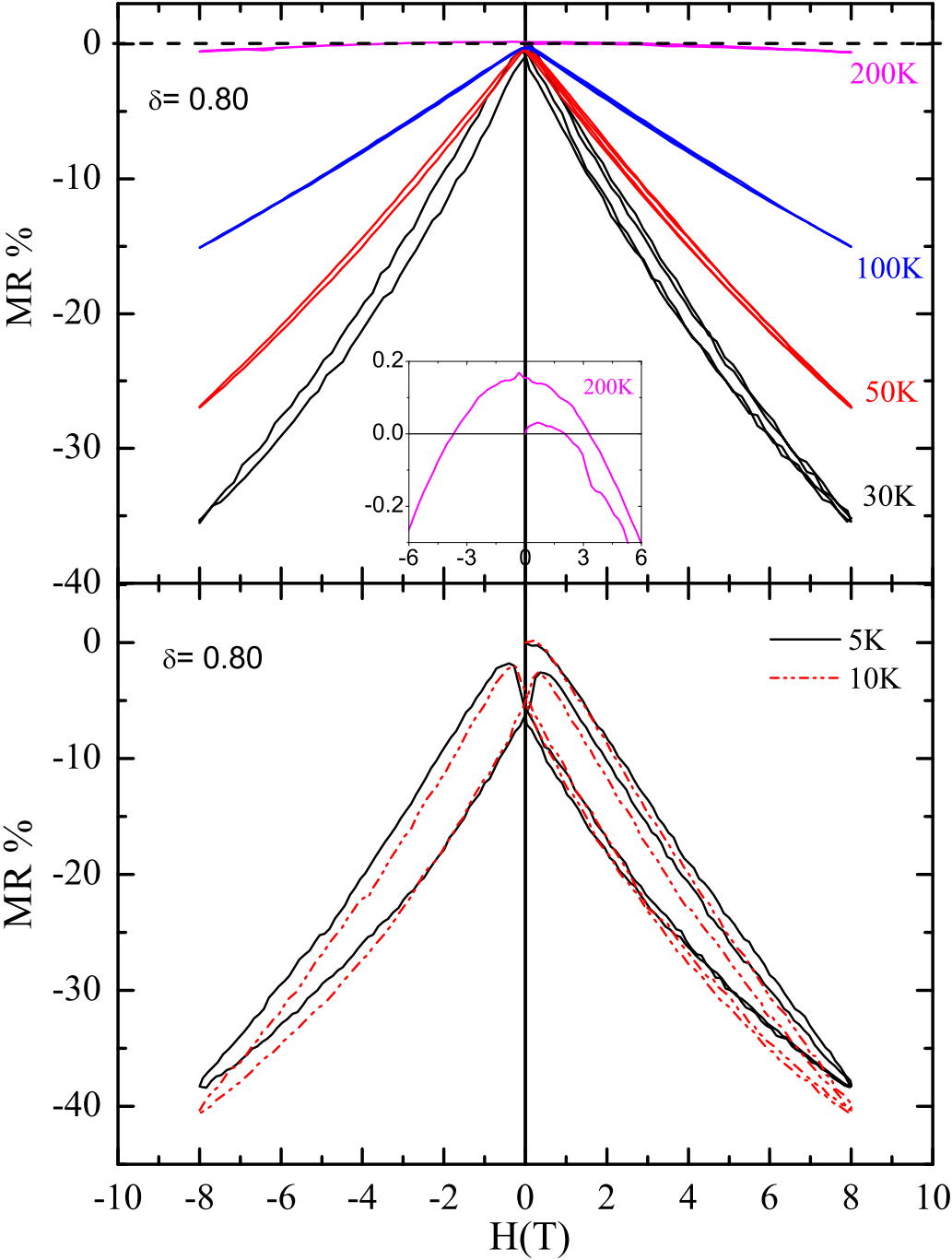}
\caption{\label{MR80} Magnetic-field-dependent isothermal magnetoresistance at various temperatures for $\delta$ = 0.80. Inset shows enlarge view of positive going MR at 200K.}
\end{figure}

Isothermal magnetoresistance (MR) for PrBaCo$_2$O$_{5.8}$ measured at various temperatures in the field range of
$\pm$ 8T are shown in Figure \ref{MR80}. MR is quite large ($\sim$ 40\% at 10K and 8T) and negative at T $\le$
100K as expected for a ferromagnetically ordered sample. An interesting feature to be noted is the
irreversibility of MR measured while ramping the magnetic field in both directions. This irreversibility of MR
gives rise to "butterfly-like" pattern below 50K. Normally such an effect is observed near magnetic phase
boundary or due to presence of competing magnetic interactions \cite{kun}. This observation augers well with the
observation of T$_{R1}$ at 50K in PrBaCo$_2$O$_{5.8}$. It may be also noted that hysteresis recorded at 5K and in
the field of $ \pm $14T fails to show saturation of magnetic moment \cite{shr}. Hence even though the compound
shows bulk ferromagnetic character below its T$_C$, there could be weak competing magnetic interactions present
due to minority Co$^{3+}$-O-Co$^{3+}$ pairs.

\begin{figure}
\centering
\includegraphics[width=\columnwidth]{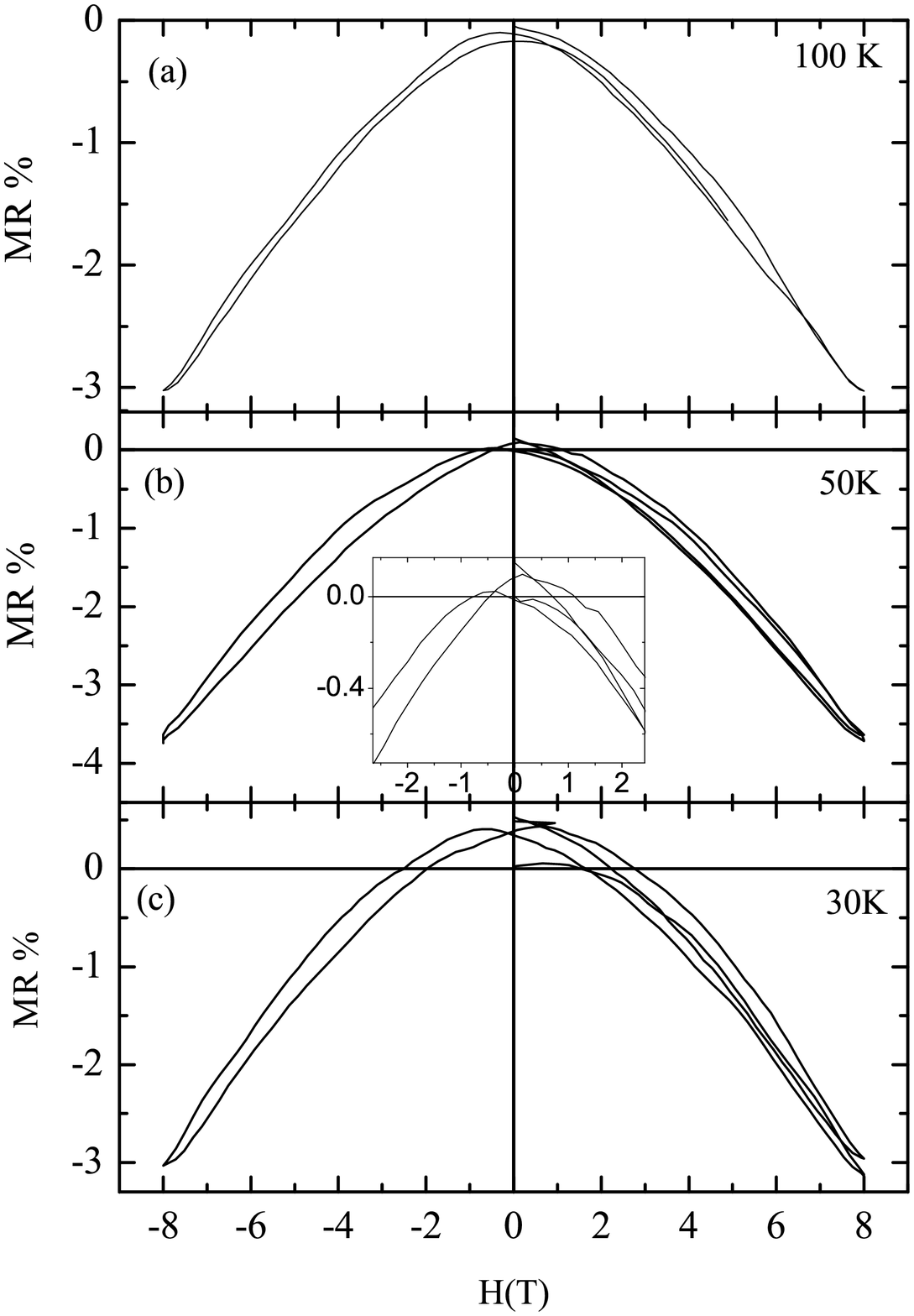}
\caption{\label{MR67} Magnetic-field-dependent isothermal magnetoresistance at various temperatures for $\delta$ = 0.67, (a) 100K, (b) 50K; Inset shows enlarged view of positive going MR at 50K, (c) 30K.}
\end{figure}

In the case of $\delta$ = 0.67 MR is negative but small ($\sim$ 3\% at 8T) and for T $\le$ 50K, exhibits positive
values at low fields (see Figure \ref{MR67}). The positive values of MR indicate antiferromagnetic ground state.
However, the crossover to negative values at low fields and butterfly-like effect hints at presence of competing
magnetic interactions or even field induced magnetization. This is in general agreement with inferences drawn
from magnetization results which show presence of antiferromagnetic correlations at higher temperatures as well
as large difference in ZFC and FC values at low temperatures. Attempts were made to measure MR at higher
temperatures but no specific trend could be made out as the values were very small.

\begin{figure}
\centering
\includegraphics[width=\columnwidth]{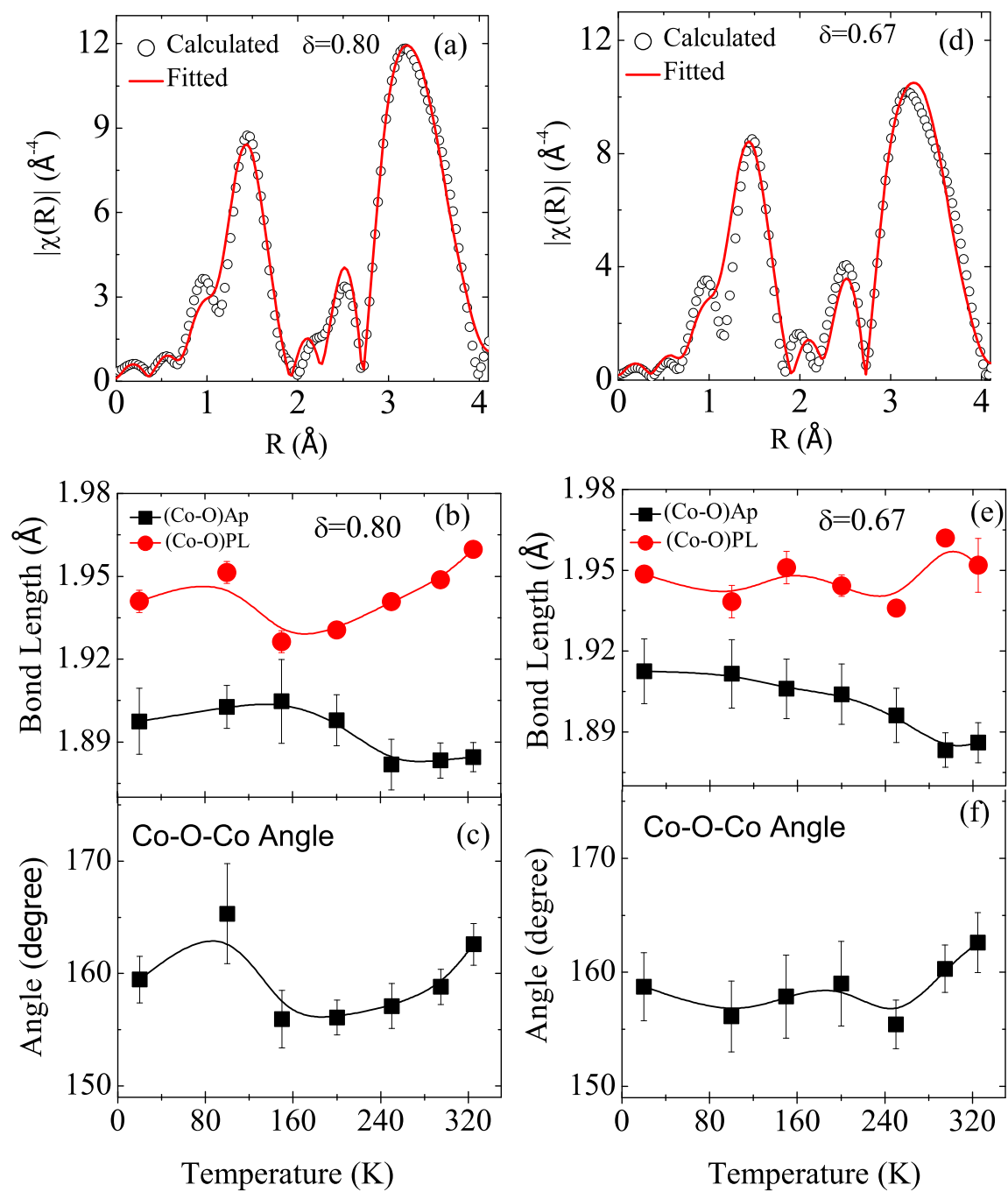}
\caption{\label{EXAFS}  (a) and (d) EXAFS $\chi$(k) data and best fit, (b) and (e) variation of average Co-O bond distances Planer and Apical (c)and (f) Co-O-Co bond angle at various temperatures for PrBaCo$_{2}$O$_{5.80}$ and PrBaCo$_{2}$O$_{5.67}$ respectively.}
\end{figure}

The presence of competing magnetic interactions, presence of short range correlations at temperatures much above
ordering temperature, irreversibility in isothermal magnetoresistance make it evidently clear that the local
structure around Co and its spin state play an important role in governing magnetic and transport properties of
these layered double perovskites. EXAFS measurements at Co K edge and at various temperatures can give an insight
of changes in local structure and spin state of Co ions. The temperature dependant EXAFS spectra recorded at
various temperatures were treated using IFEFFIT software \cite{matt}. The absorption data was first processed
using Athena to extract the EXAFS signal $\chi(k)$ from $\mu$ v/s E data \cite{ravel}. The $\chi(k)$ v/s k data
was Fourier transformed (FT) and fitted with theoretically calculated EXAFS signal for the known crystal
structure in Artemis \cite{ravel1}. The theoretical standards were calculated using FEFF 6.01 \cite{rehr} and
lattice parameters from Table\ref{lattice}. Figure \ref{EXAFS} (a) and (d) presents magnitude of FT of $k^3$
weighted $\chi(k)$ data for the two compounds $\delta$ = 0.8 and 0.67 respectively. EXAFS spectra in the range
1-4 \AA~ were fitted with Co-O, Co-Co and Co-Pr/Ba single scattering and Co-O-Co multiple scattering
correlations. The fitting parameters were defined using relations based on lattice parameters to reduce
interdependency and number of variables. The solid lines in Figure \ref{EXAFS} (a) and (d) respectively depict
the best fit obtained to experimental data at 20K. The variations of Co-O bond distances and Co-O-Co bond angle
obtained from fitting are presented in Figure \ref{EXAFS} (b) and (c) for $\delta$ = 0.80 and (e) and (f) for
$\delta$ = 0.67 respectively.

In the case of $\delta$ = 0.8, it can be seen that the Co-O planar bond length (Figure \ref{EXAFS}(b)) decreases
steadily with temperature followed by an increase around 150K which coincides with the T$_C$ of the compound. A
similar behavior is also noted for the planar Co-O-Co bond angle (Figure \ref{EXAFS}(c)). On the other hand, the
Co-O apical bond distance exhibits a step like behavior with an increase around 240K. This elongation of CoO
polyhedra along $c$-axis and the concomitant shrinkage in Co-O planar bond length and the Co-O-Co bond angle
corresponds to the tilting of CoO polyhedra towards each other. Further, at lower temperatures ($\sim$ 150K) the
polyhedra rotate about c axis resulting in increase of Co-O planar bond length and the planar Co-O-Co bond angle.

A similar variation of bond distances can be seen in the case of $\delta$ = 0.67. The elongation along c-axis and
tilting of Co-O polyhedra can be seen to happen around 300K (Figure \ref{EXAFS}(e), (f)). At lower temperatures,
the apical Co-O bond distance is found to remain nearly constant while the planar Co-O bond distance and Co-O-Co
bond angle exhibit identical variation with temperature. In particular, the planar Co-O bond distance and bond
angle show two minima centred at about 250K and 100K and a maxima at around 150K. Interestingly these
temperatures respectively correspond to onset of magnetic correlations as seen from Curie-Weiss analysis of
susceptibility, and the antiferromagnetic ordering temperatures, T$_{N1}$ and T$_{N2}$ of the compound.

\section{Discussion}
The two compounds under study PrBaCo$_2$O$_{5+\delta}$ with $\delta $= 0.67 and 0.80 crystallize in tetragonal
unit cell with P4/mmm structure. However, they exhibit contrasting magnetic behavior. While PrBaCo$_2$O$_{5.8}$
orders ferromagnetically below 147K, PrBaCo$_2$O$_{5.67}$ has an antiferromagnetic ground state. In the case of
PrBaCo$_2$O$_{5.80}$, charge balance demands 30$\%$ Co ions to be in tetravalent state while the rest 70$\%$ to
be in trivalent state.
The value of $\mu_{eff}$ obtained from Curie-Weiss analysis of susceptibility and its agreement with that
reported for PrBaCo$_2$O$_{5.75}$ \cite{fr} indicates that Co$ ^{4+} $ ion are in LS state while Co$ ^{3+} $ ions
are in HS/IS state. Therefore the observed FM ordering can be ascribed to double exchange (DE) interaction
between LS Co$^{4+}$ ($t_{2g}^5,e_g^0$) and HS/IS Co$^{3+}$ ($t_{2g}^5,e_g^1$ or $t_{2g}^4,e_g^2$) ions. These DE
interactions are facilitated by tilting and the rotation of Co$^{3+}$O and Co$^{4+}$O polyhedra towards each
other about the $c$ axis. This increases the Co(3d) - O(2p) hybridization leading to a ferromagnetic order. The
change in hybridization also promotes hopping conduction behavior resulting in lower values of resistivity.
Mott's VRH law is best obeyed in the temperature region between 200K and 50K. Deviations from linearity can be
noticed below and above this temperature range. MR also exhibits be irreversibility at T$ \leq $ 50K and positive
values at low fields at T = 200K. It may be noted that T = 200K corresponds approximately to the temperature at
which Co-O polyhedra tilt towards each other while the weak AFM interactions noticed below 50K could be implied
to be due to superexchange (SE) interactions between Co$^{3+}$ ions as per Goodenough-Kanamori rule.

For PrBaCo$_2$O$_{5.67}$, the properties are found to be very different. With the decrease in oxygen content
though the percentage of Co in square pyramidal coordination  and those in trivalent state increases the value of
$\mu_{eff}$ obtained from Curie Weiss fit to susceptibility is nearly the same as that in PrBaCo$_2$O$_{5.80}$.
This implies that Co$^{3+}$ continue to be in HS/IS state and the Co$^{4+}$ ions in LS state. Therefore
antiferromagnetic superexchange interaction can be expected between neighboring CoO polyhedra containing
trivalent cobalt ions while the Co$^{3+}$-O-Co$^{4+}$ pairs will interact via double exchange mechanism. EXAFS
studies indicate that the antiferromagnetic SE interactions are a result of the outward rotation of Co$^{3}$O
polyhedra about the $c$-axis. The presence of competing ferromagnetic interactions arising due to DE interactions
can give rise to canting of Co spins resulting in an increase in magnetization at about 150K.  The
antiferromagnetic order localizes the charge carriers leading to a much higher resistivity as compared to
$\delta$ = 0.8 compound and a much lower MR even near the AFM transition.

\section{Conclusion}
In summary, it can be concluded that the local structure and the changes therein play an important role for
magnetic and transport properties of layered perovskites. In the case of $\delta$ = 0.80 the inward rotation and
tilting of the neighboring Co$^{3+}$O and Co$^{4+}$O polyhedra about the c-axis facilitates DE mechanism between
LS Co$^{4+}$ and HS/IS Co$^{3+}$ ions leading to ferromagnetic order. While in case of $\delta$ = 0.67 the
tilting and outward rotation of the neighboring Co$^{3+}$O polyhedra away from each other supports Co$^{3+}$ - O-
Co$^{3+}$ superexchange mediated antiferromagnetic ordering.

\ack{Acknowledgements} KRP and SG would like to acknowledge the travel support from Department of Science and
Technology, Govt. of India under the Utilization of International Synchrotron Radiation and Neutron Scattering
facilities and financial assistance from Council for Scientific and Industrial Research, New Delhi under the
project EMR-II/1099. The experiments at PF were performed under the Proposal No. 2009G215. SG acknowledges the
travel assistance and local hospitality extended to her by Centre Director, UGC-DAE Consortium, Indore.

\end{document}